\documentclass[manuscript]{aastex}

\shorttitle{\emph{Chandra} Proper Motion of PSR J0357+3205}
\shortauthors{De Luca et al.}

\begin{document}

\title{PSR J0357+3205: a fast moving pulsar with a very unusual X-ray trail.}

\author{A.~De Luca\altaffilmark{1,2}, R.~P.~Mignani\altaffilmark{1,3,4},
M.~Marelli\altaffilmark{1}, D.~Salvetti\altaffilmark{1,5},
N.~Sartore\altaffilmark{1}, A.~Belfiore\altaffilmark{6}, 
P.~Saz Parkinson\altaffilmark{6}, P.A.~Caraveo\altaffilmark{1,2}, 
G.F.~Bignami\altaffilmark{7,1}}

\altaffiltext{1}{INAF - Istituto di Astrofisica Spaziale e Fisica Cosmica Milano, Via 
E.~Bassini 15, 20133 Milano, Italy}
\altaffiltext{2}{Istituto Nazionale di Fisica Nucleare, Sezione di Pavia, Via
  Bassi 6, 27100 Pavia, Italy}
\altaffiltext{3}{Mullard Space Science Laboratory, University College London, 
  Holmbury St. Mary, Dorking, Surrey, RH5 6NT, UK}
\altaffiltext{4}{Kepler Institute of Astronomy, University of Zielona G\'ora, 
  Lubuska 2, 65-265, Zielona G\'ora, Poland}
\altaffiltext{5}{Universit\'a degli Studi di Pavia, Dipartimento di Fisica, Via
  Bassi 6, 27100 Pavia, Italy}
\altaffiltext{6}{Santa Cruz Institute for Particle Physics, Department of
  Physics, University of California at Santa Cruz, Santa Cruz, CA 95064, USA}
\altaffiltext{7}{Istituto Universitario di Studi Superiori di Pavia,
 Piazza della Vittoria n.15, 27100 Pavia, Italy}
\email{deluca@iasf-milano.inaf.it}

\begin{abstract}
The middle-aged PSR J0357+3205 is a nearby, radio-quiet, bright $\gamma$-ray 
pulsar discovered by the Fermi mission. Our previous \emph{Chandra} observation
revealed a huge, very peculiar structure of diffuse X-ray
emission, originating at the pulsar position and extending for $>\,9'$ on the plane of the
sky. To better understand the nature of such a nebula, we have studied the 
proper motion of the parent pulsar. We performed relative astrometry 
on \emph{Chandra} images of the field spanning a time baseline of 2.2 yr, 
unveiling a significant angular displacement of the pulsar counterpart,
corresponding to a proper motion of $0\farcs165\pm0\farcs030$ yr$^{-1}$. At a
distance of $\sim500$ pc, the space velocity of the pulsar would be of
$\sim390$ km s$^{-1}$ assuming no inclination with respect to the plane of the
sky. The direction of the pulsar proper motion is perfectly aligned
with the main axis of the X-ray nebula, pointing to a physical, yet elusive
link between the nebula and the pulsar space velocity.
No optical emission in the H$_{\alpha}$ line is seen in a deep image
collected at the Gemini telescope, which implies that the interstellar
medium into which the pulsar is moving is fully ionized.
\end{abstract}


\section{Introduction}
The Large Area Telescope onboard the \emph{Fermi} satellite 
\citep[\emph{Fermi}-LAT,][]{atwood09} has opened a new era
for pulsar astronomy, by detecting $\gamma-$ray
pulsations (at E$>$100 MeV) from more than 120 
pulsars\footnote{See \url{https://confluence.slac.stanford.edu/display/GLAMCOG/Public+List+of+LAT-Detected+Gamma-Ray+Pulsars/}},
about 30\% of which are not detected at radio wavelengths. 
The middle-aged PSR J0357+3205 (characteristic age $\tau_C\sim0.54$ Myr)
is one of the most interesting radio-quiet pulsars
discovered in blind periodicity
searches in \emph{Fermi}-LAT data \citep{abdo09a}. Its high $\gamma$-ray flux
\citep[it is included in the \emph{Fermi}-LAT bright source list,][]{abdo09b},
low spin-down luminosity ($\dot{E}_{rot}=5\times10^{33}$ erg s$^{-1}$) 
and off-plane position (Galactic latitude $b\sim-16^{\circ}$)
point to a small distance of about 500 pc.  
We investigated the field of PSR J0357+3205 with a joint
X-ray  and optical  program with \emph{Chandra} and the NOAO 
Mayall 4m telescope at Kitt Peak. This allowed us to identify the soft X-ray
counterpart of the pulsar as an unremarkable source, looking 
fainter (and colder) than other well known middle-aged pulsars \citep{deluca11}.
More interestingly, our deep \emph {Chandra} observation unveiled the existence
of a very large, elongated feature of diffuse X-ray emission,
apparently originating at the pulsar position and extending for
more than $9'$ (corresponding
to $\sim1.3$ pc at the distance of 500 pc, assuming no inclination
with respect to the plane of the sky), with a hard spectrum 
consistent both with a power law and with a hot thermal 
bremsstrahlung.

Elongated ``tails'' of diffuse emission
have been  associated with several rotation-powered pulsars
\citep[e.g.][]{kargaltsev08} and explained 
as bow-shock pulsar wind nebulae \citep[see][for a review]{gaensler06}, 
where their
elongated morphology is ``velocity-driven''. Indeed, if the
pulsar moves supersonically through the interstellar medium, 
the termination shock of the pulsar wind assumes
a ``bullet'' morphology, due to ram pressure. Particles
accelerated at the shock 
emit
synchrotron radiation
and cool down, 
confined by ram pressure in an elongated region
aligned with the pulsar space velocity.
However, explaining the nature of the nebula associated
with PSR J0357+3205 turned out to be very challenging.
As discussed by \citet{deluca11},
the standard picture cannot apply here since the 
morphology is very different from the ``cometary'' shape
which characterizes all other X-ray bow-shock nebulae.
There is no emission in the surroundings of the pulsar,
where the brightest portion (the termination shock)
should be -- indeed, the surface brightness grows as a function 
of the distance from PSR J0357+3205. Moreover, there 
are no evidences for spectral evolution as a function of 
the position, at odds with expectations for a population of particles
injected at the shock and cooling via synchrotron radiation.

Other pictures could be explored. For instance, PSR B2224+65,
the fast moving pulsar powering the well known ``Guitar nebula''
seen in H$_{\alpha}$ \citep{cordes93},
displays an elongated X-ray feature which is reminiscent
of the one of our target and cannot be a bow-shock nebula 
because it is misaligned by $\sim118^{\circ}$ with respect
to the direction of the proper motion \citep{hui07}. Thus, the 
possibility of a ballistic jet (similar to Active Galactic Nuclei),
or the hypothesis of a nebula confined by a pre-existing, 
large scale magnetic field in the interstellar medium
have been proposed \citep{bandiera08,johnson10,hui12}.

Indeed, a crucial piece of information in order to understand the
physics of the huge elongated feature associated with PSR J0357+3205
is the direction of the pulsar proper motion. Detecting a pulsar 
angular displacement aligned with the nebula's main axis would link
the morphology of the diffuse structure to the pulsar velocity.
Conversely, if it were misaligned, the case of PSR J0357+3205 would become very similar
to the one of PSR B2224+65 and would require a different explanation
for the nature of the nebula.

Usually, a pulsar proper motion is measured in the radio band,
or, more rarely, in the optical domani.
Unfortunately, our target is radio-quiet and has no optical counterpart; 
moreover, timing analysis of gamma-ray photons is not particularly sensitive 
to the proper motion \citep[positional accuracy based on 5 yr of 
\emph{Fermi}-LAT timing is estimated to be $\sim2''$,][]{ray11}. 
The only way to search 
for a possible proper motion rests on the comparison of multi-epoch, 
high-resolution X-ray images. To this aim, we have obtained a multi-cycle
observing campaign with Chandra, consisting of two observations to be
performed at the end of 2011 and at the end of 2013.
We will report here on 
the first observation of our program, as well as
on a very recent observation of the field 
in the 
H$_{\alpha}$ band performed with GMOS instrument at the Gemini North telescope. 
Indeed, pulsars moving supersonically into warm interstellar gas can generate
optical emission in the H$_{\alpha}$ line, due to collisional
excitation of neutral hydrogen 
and charge exchange occurring at (and behind) the forward shock
\citep[see e.g.][]{cordes93},
yielding 
a limb-brightened, arc-shaped bow-shock nebula,
located at the apex of the forward shock, in the direction of the proper
motion. 

\section{Measurement of the pulsar proper motion}
The superb angular resolution of the \emph{Chandra} optics 
makes it possible to measure
tiny angular displacements of an X-ray source 
by performing relative astrometry on multi-epoch images.
Indeed, such an approach has already been used to measure
the proper motion of a few isolated neutron stars. See e.g.
our investigation for the case of SGR 1900+14
\citep{deluca09}, as well as the work by \citet{motch07},
\citet{motch08}, \citet{kaplan09}, \citet{becker12}, \citet{vanetten12}.

\subsection{Chandra observations and data reduction}
Our new observation of PSR J0357+3205 with \emph{Chandra} was performed
on 2011, December 24 (Obs.Id. 14007, 29.4 ks exposure time -- hereafter 
tagged as ``2011'').
Previous observations were performed on 2009 October 25 (obs. id. 12008, 29.5
ks -- hereafter ``2009a'') 
and on 2009 October 26 (obs. id. 11239, 47.1 ks -- hereafter ``2009b'').
All data were collected using the Advanced CCD Imaging Spectrometer (ACIS)
instrument in Timed Exposure mode with the VFAINT telemetry mode. 
We retrieved ``Level 1'' data from the \emph{Chandra} Science Archive and reprocessed
them with the {\tt chandra\_repro}\footnote{{\texttt
    http://cxc.harvard.edu/ciao/ahelp/chandra\_repro.html}} 
script of the \emph{Chandra} Interactive Analysis 
of Observation Software (CIAO v4.4)\footnote{{\texttt http://cxc.harvard.edu/ciao/index.html}}. 

For each observation, we generated an
image in the 0.3--8 keV energy range using the original ACIS pixel size 
($0\farcs492$/pixel). We performed a source detection using the 
{\tt wavdetect}\footnote{{\texttt http://cxc.harvard.edu/ciao/threads/wavdetect/}} task
with wavelet scales ranging from 1 to 16 pixels with a $\sqrt{2}$ step size, 
setting a detection threshold of $10^{-6}$. 
In all observations, the target was imaged close to the aimpoint, 
on the backside-illuminated chip S3 of the ACIS-S array.

We cross-correlated the resulting source lists using a correlation radius of
3$''$ and  we extracted a catalogue 
of common sources for each pair of observations. 
In view of the density of sources in each image, 
the possibility of a chance alignement of two false detections is $<10^{-5}$.
As a further step, we selected sources within $4'$ of the aimpoint
since the telescope point spread function deteriorates as a function of 
offaxis angle, hampering source localization accuracy \citep[see discussion in][]{deluca09}.
Such an exercise yielded 12 common sources (2011 vs. 2009a), 10 sources (2011 vs. 2009b)
and 16 sources (2009a vs. 2009b), in addition to the target pulsar.
The uncertainty on the source localization on each image depends
on the signal to noise as well as on the offaxis angle and 
ranges from $\sim0.08$ to $\sim0.7$ pixel per coordinate.

\subsection{Relative astrometry}
The positions of the selected, common sources 
(excluding the pulsar counterpart) were adopted as a reference grid
to perform relative astrometry. We used the ACIS SKY reference 
system\footnote{{\texttt http://cxc.harvard.edu/ciao/ahelp/coords.html}}
(pixel coordinates with axes aligned along Right Ascension and Declination).
Taking into account the corresponding uncertainties,
we computed the best geometric transformation 
needed to superimpose the reference frames of two
images collected at different epochs.

We superimposed the most recent 2011 data 
to first-epoch 2009a and (separately) 2009b data in order to measure 
the possible pulsar displacement over
a baseline of $\sim2.2$ yr. We also superimposed 2009a data to 2009b 
data (no displacement is expected on a 1-day baseline)
in order to check for systematics affecting our analysis.
A simple translation 
yielded a good superposition in all cases (see Table 1). 
The uncertainty on the frame 
registration turned out to be smaller than 50 mas per coordinate.
Adding a further free parameter to the transformation (a rotation
angle) did not result in a statistically compelling improvement
of the fit. 

\begin{table}
\begin{center}
\caption{Results of X-ray relative astrometry}
\begin{tabular}{l|c|c|c}
\tableline \tableline
 & 2011 vs. 2009a & 2011 vs. 2009b & 2009a vs. 2009b \\ 
\tableline \tableline
Time baseline & 2.16 yr & 2.16 yr & 1 day \\
\tableline
Number of ref.srcs & 11\tablenotemark{a} & 10 & 16 \\
\tableline
uncertainty on X$_{shift}$ (pixels) & 0.09 & 0.08 & 0.06 \\
$\chi^2$ (dof) & 13.6 (10) & 15.6 (9) & 7.8 (15) \\
\tableline
uncertainty on Y$_{shift}$ (pixels) & 0.08 & 0.07 & 0.06 \\
$\chi^2$ (d.o.f.)  & 8.1 (10) & 13.0 (9) & 13.8 (15) \\
\tableline
PSR X displacement (pixels) & $0.53\pm0.12$ & $0.50\pm0.11$ & $0.10\pm0.10$ \\
PSR Y displacement (pixels) & $0.54\pm0.11$ & $0.50\pm0.10$ & $0.04\pm0.10$ \\
\tableline \tableline
\tablenotetext{a}{A reference source yielding high residuals was rejected.}
\end{tabular}
\end{center}
\end{table}

We applied the best-fit transformation to the 
coordinates of the pulsar counterpart and we computed its 
displacement between different epochs.
The error budget for the overall pulsar displacement includes 
the uncertainty on the pulsar position in each image as well as
the uncertainty on the multi-epoch frame registration. Results are 
shown in Table 1. 
Displacement of the pulsar as well as residuals on the 
positions of the reference sources after frame registration are also 
shown in Figure 1.
A significant and consistent displacement of 
the pulsar is apparent both in the 2011 vs. 2009a 
and in the 2011 vs. 2009b comparison, while 
no displacement is seen  for the pulsar
in the 2009a vs. 2009b one.

By combining the two measurements, we obtain a pulsar proper motion
$\mu_{\alpha}cos({\delta})=117\pm20$ mas yr$^{-1}$ 
and $\mu_{\delta}=115\pm20$  mas yr$^{-1}$. 
This corresponds to a total proper motion of 165$\pm$30 mas yr$^{-1}$,
translating to a (projected) space velocity of $\sim390$d$_{500}$ km s$^{-1}$
(where d$_{500}$ is the distance to the pulsar in units of 500 pc). 
The position angle of the proper motion is $314^{\circ}\pm10^{\circ}$ (North to East)
(see Figure 2). The proper motion direction is very well aligned with the
main axis of the nebula, which is indeed an  ``X-ray trail''.

\section{Gemini H$_{\alpha}$ observations.}
We observed the field of PSR J0357+3205 in the H$_{\alpha}$ band on September 23,
2012 with the Gemini-North telescope on the Mauna Kea Observatory. 
We used the Gemini Multi-Object Spectrograph (GMOS) in its imaging mode
equipped with the interim upgrade e2v deep depletion (DD) detector, 
with enhanced response in the blue and red arm of the spectrum with respect
 to the original EEV detector. The e2v DD detector is an array of three chips, 
with an unbinned pixel size of 0\farcs072 8 and covers an unvignetted  
field--of--view of $5\farcm5 \times 5\farcm5$. We used the Ha\_G0310 filter 
($\lambda=656$ nm; $\Delta \lambda=7$ nm), centered on the H$_{\alpha}$ rest 
wavelength. The e2v DD detector was set in the $2\times2$ binning mode and
read through the standard slow read-out/low gain mode with the six
amplifiers. Observations were performed with an average airmass of 1.24, 
image quality of $\sim 0\farcs8$, and grey time. Two sets of six 500 s
exposures were obtained, for a total integration time of 6000 s. 
Exposures were dithered in steps of $\pm$ 5 pixels to achieve a better 
signal--to--noise (S/N).

We processed the GMOS images using the dedicated  {\sc gmos}  image reduction 
package available in {\sc IRAF}. After  downloading the closest 
bias and sky flat-field frames from the {\em Gemini}  science 
archive\footnote{{\texttt http://cadcwww.dao.nrc.ca/gsa/}}, we used the  
tasks {\tt gbias} and {\tt giflat} to process and combine the bias and  
flat-field frames, respectively.  We then reduced the single science frames 
using the task {\tt gireduce} for bias subtraction, overscan correction, 
image trimming and flat-field normalisation.  From  the  reduced science 
images, we produced a mosaic of the three GMOS chips using the task 
{\tt  gmosaic} and we used the task  {\tt imcoadd} to average-stack the 
reduced image mosaics and filter out cosmic ray hits. We computed the 
astrometry calibration using the 
{\em wcstools}\footnote{http://tdc-www.harvard.edu/wcstools/} suite of 
programs, matching the sky coordinates of stars selected from the Two Micron 
All Sky Survey (2MASS) All-Sky Catalog of Point Sources \citep{skr06} 
with their pixel coordinates computed by {\em Sextractor} 
\citep{bs96}.  After iterating the matching process applying 
a $\sigma$-clipping selection to filter out obvious mismatches, 
high-proper motion stars, and false detections, we obtained mean residuals
of $\sim 0\farcs2$ in the radial direction, using up to 30 bright, 
but non-saturated,  2MASS stars. To this value we added in quadrature 
the uncertainty $\sigma_{tr}$ = 0\farcs08 of the image registration  
on the  2MASS reference frame. This is given by 
$\sigma_{tr}$=$\sqrt{n/N_{S}}\sigma_{\rm S}$ \citep[e.g.][]{lat97}, 
where $N_{S}$ is the number of stars used to compute the astrometric 
solution, $n$=5 is the number of free parameters in the sky--to--image 
transformation model, $\sigma_{\rm S} \sim 0\farcs2$ is the mean absolute 
position error of  2MASS \citep{skr06} for stars in the magnitude 
range  $15.5 \le K \le 13$. After accounting for the 0\farcs015  uncertainty 
on the link of 2MASS to the International Celestial Reference Frame  
\citep{skr06}, we evaluated with the overall accuracy on our absolute
astrometry
to be of $\sim$0\farcs22.

Unfortunately, no observations of spectro-photometric standards stars were 
available for the flux calibration of our  H$_{\alpha}$ image. 
Thus, we cross-correlated instrumental magnitudes of more than
150 non-saturated stars in the Gemini image to their R$_F$ magnitudes
listed in the Guide Star Catalogue v2.3.2 \citep[GSC2,][]{lasker08},
which yielded a rather good fit with a r.m.s. of 0.16 mag.
To assess the GSC2 flux zeropoint, we cross-correlated 
the GSC2 $R_F$ magnitude of more than 500 stars
with their R magnitude as tabulated in the Stetson
Standard photometric star 
archive\footnote{\url{http://www3.cadc-ccda.hia-iha.nrc-cnrc.gc.ca/community/STETSON/archive/}}. 
We evaluated
the transformation as $R_F$=0.97R-0.32 ($\chi^2=617$, 520 d.o.f.).
Then, we computed the H$_{\alpha}$ fluxes of the 150 selected stars
on our Gemini image
using specific fluxes corresponding to their $R_F$ magnitudes
and assuming a flat spectrum within the R filter bandpass.
This yielded a flux calibration of the H$_{\alpha}$ image
with an uncertainty of $\sim0.2$ mag.

No structures of diffuse emission unambiguously related to the 
fast motion of PSR J0357+3205
can be discerned on our Gemini image -- 
in particular, no arc-shaped ``bow-shock'' nebula
is seen at the expected position in the direction of the pulsar 
proper motion (see Figure 3). We measured the background properties in the region 
surrounding the pulsar position. Considering a region of 
10 square arcsec as a reference and taking into 
account the uncertainty in 
the flux calibration of our H$_{\alpha}$ image, we computed
the $5\sigma$ upper limit to the surface brightness of an unseen
bow-shock nebula to be of $5\times10^{-18}$ erg cm$^{-2}$ s$^{-1}$ 
arcsec$^{-2}$. 

\section{Discussion}
Relative astrometry on our multi-epoch \emph{Chandra} images 
unveiled a significant proper motion
of $165\pm30$ mas yr$^{-1}$ for PSR J0357+3205,
corresponding to a velocity of $\sim390$d$_{500}$ km s$^{-1}$,
along a direction almost coincident
with the main axis of the elongated X-ray nebula.

Our result points to a direct link
between the nebula morphology (and physics)
and the space velocity of the pulsar -- supersonic
for any reasonable condition of the interstellar medium
\citep[typical values are of $\sim1$, $\sim10$ and $\sim100$
km s$^{-1}$ for the cold, warm and hot components, respectively, see e.g.][]{kulkarni88}.
As already mentioned, ``velocity-driven'' pulsar wind
nebulae are a well-known astrophysical reality. 
Could the X-ray trail of PSR J0357+3205 be explained as 
a bow-shock PWN? As discussed by \citet{deluca11},
the lack of any evidence of the pulsar wind termination shock
could be due to its confusion with the pulsar emission.
Assuming a velocity of 390d$_{500}$ km s$^{-1}$  
and no inclination with respect to the plane of the sky,
the expected angular separation between the pulsar and the
forward and backward termination shock is $0\farcs3n_{1,ISM}^{-1/2}$ 
and $2''n_{1,ISM}^{-1/2}$,
respectively \citep{deluca11}, where $n_{1,ISM}$ is the
density of the interstellar medium in units of 1 particle 
cm$^{-1}$. Thus, high inclination of the pulsar velocity
and/or high density of the ISM, and/or a distance larger 
than expected should be assumed; any 
of these hypotheses could be possible (though 
rather unlikely).
In any case, other peculiarities of the trail phenomenology
cannot be accounted for in the standard scenario of 
synchrotron emission from shocked pulsar wind.
One should explain why the maximum luminosity
is generated at a distance as large as $\sim3$ light years 
from the parent neutron star. This would require 
an increase of the density of the radiating particles 
on the same distance scale (possibly due to
deceleration of the bulk flow?), or an increase of the magnetic 
field intensity, or a change of the angle between the
magnetic field and the particle flow. 
Moreover, the lack of 
any spectral steepening across the trail \citep{deluca11} 
clashes with the expected, severe synchrotron energy losses
of particles radiating in the keV range.
Thus, either some very peculiar phenomenon is occurring in the 
magnetized pulsar particle outflow 
(local particle re-acceleration? large-scale  
organization of the flow magnetic field?) 
or the nature of the
nebula is different (but linked to the pulsar velocity).
For instance, \citet{marelli12} propose that the trail
could be due to thermal emission by the ISM heated 
by the shock driven by the fast moving pulsar.

The measure of the  pulsar space velocity can be used, 
together with the upper limit to the surface brightness 
of any undetected diffuse structure in our Gemini H$_{\alpha}$ image,
to constrain the conditions of the medium in which 
PSR J0357+3205 is moving. 
The expected flux of
the narrow-line component of
a pulsar H$_{\alpha}$ bow-shock nebula depends
on the pulsar spin-down luminosity $\dot{E}_{rot}$, 
the pulsar space velocity v$_{psr}$, the distance to the source D$_{psr}$, 
and the neutral fraction of the medium
X$_{ISM}$. 
Using the scaling law  proposed by \citet{cordes93} and \citet{chatterjee02},
we estimated that
for any reasonable value of the distance to PSR J0357+3205 (ranging from
$\sim200$ pc, as suggested by the non negligible X-ray absorbing column, to $\sim900$ pc,
where $\gamma$-ray luminosity would exceed the spin-down luminosity)
and in a broad range of possible space velocities for the pulsar
(assumining the inclination angle with respect to the 
plane of the sky  in the $\pm75^{\circ}$ range), 
the lack of detection of a H$\alpha$ nebula can only be
ascribed to a neutral fraction X$_{ISM}<$0.01, i.e.
the gas surrounding the pulsar
is fully ionized. 
Thus, PSR J0357+3205 could be
traveling across a region filled with ISM in the hot phase.
Some contribution to the ionization of the medium 
from the pulsar itself is also possible, although
its thermal emission is not particularly prominent
\citep{deluca11}.


PSR J0357+3205 is located well away from the Galactic plane
(b$\sim-16^{\circ}$). In view of this, it is interesting 
to note that the direction of the pulsar proper motion
is almost parallel to the Galactic plane
(it has an inclination of $\sim2^{\circ}$ towards the Plane). 
As done in \citet{mignani12}, we used the proper motion information 
to extrapolate back in time the trajectory of the pulsar in the 
Galactic potential to find possible associations with open clusters 
and OB associations \citep{dias02,melnik09,dezeeuw99}, 
leaving the pulsar radial velocity as a free parameter 
and assuming a distance of 500 pc and an age of 0.54 Myr. 
However, we found no association with clusters or OB associations 
closer than $\sim2$ kpc. In any case, such an exercise
showed us that if PSR J0357+3205 has a radial velocity of order
600 km s$^{-1}$ (or greater) in the direction away from the Solar 
System, the pulsar could have been born on the Galactic plane
\citep[assuming a scale height of 60 pc, typical for massive stars, see e.g.][]{maiz01}. 
Thus, the hypothesis of a runaway high-mass star 
as the progenitor of PSR J0357+3205, suggested by \citet{deluca11},
is not required in a range of possible orbital solutions.

\section{Conclusions}
The radio-quiet PSR J0357+3205 is traveling very fast ($\sim390$ km s$^{-1}$) into 
a fully ionized interstellar medium. Its proper motion is perfectly
aligned with its very elongated X-ray nebula, pointing
to a physical link between the nebula and the space velocity of the neutron
star. However, the standard scenario of a bow-shock synchrotron-emitting
nebula does not explain the phenomenology of our target.
Deep radio polarization images could help explain 
the nature of this PWN, by tracing the underlying large-scale magnetic 
field configuration. Unveiling a population of energetic, nearby pulsars 
missed by radio surveys, the \emph{Fermi} mission is opening new avenues to detect 
unusual PWNe which could point to different ways to deposit pulsar rotational 
energy into the interstellar medium.

\acknowledgments
RPM thanks Andrew Gosling for his support during the preparation of the Gemini
observations. The research leading to these results has received funding from 
the European Commission Seventh Framework Programme (FP7/2007-2013) under
grant agreement n. 267251. This research used the facilities of the Canadian 
Astronomy Data Centre operated by the National Research Council of Canada with 
the support of the Canadian Space Agency.

{\it Facilities:} \facility{CXO (ACIS)} \facility{Gemini:Gillett}.



\clearpage

\begin{figure}
\epsscale{.40}
\plotone{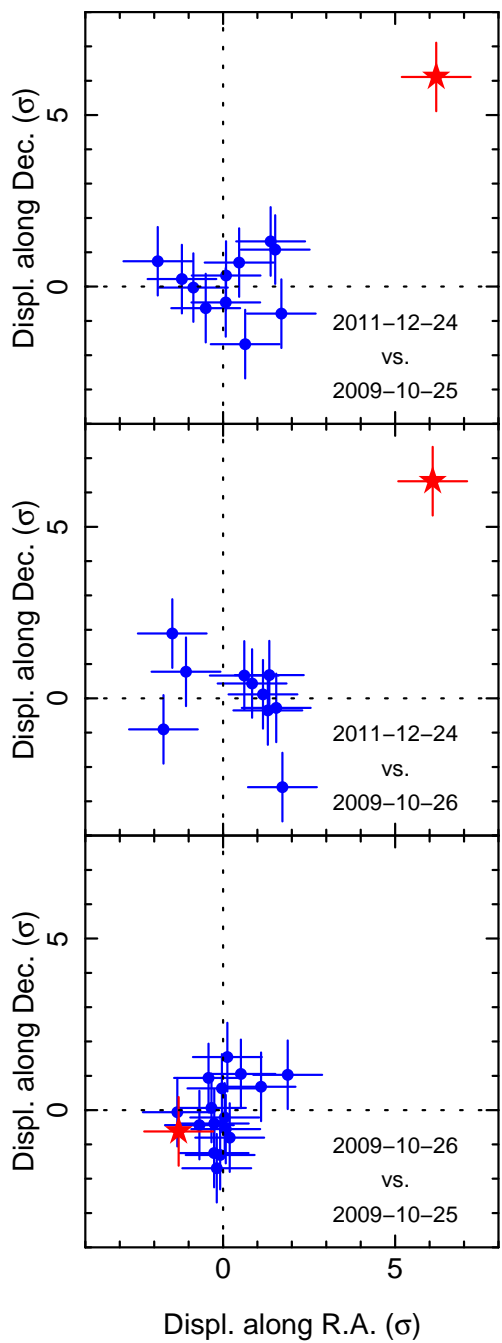}
\caption{Results of relative astrometry. Two independent, first-epoch images
(2009 October 25 and 2009 October 26) are compared to a second-epoch image
(2011 December 24) in the upper and middle panels, respectively. Coordinate
residuals after image superposition are shown for reference sources (blue
dots) as well as for the pulsar counterpart (red star), in units of
statistical errors. The displacement of the pulsar is apparent. The bottom box
displays the comparison of the two first-epoch images showing, as expected, 
no displacements for any source.}
\end{figure}

\clearpage

\begin{figure}
\epsscale{.80}
\plotone{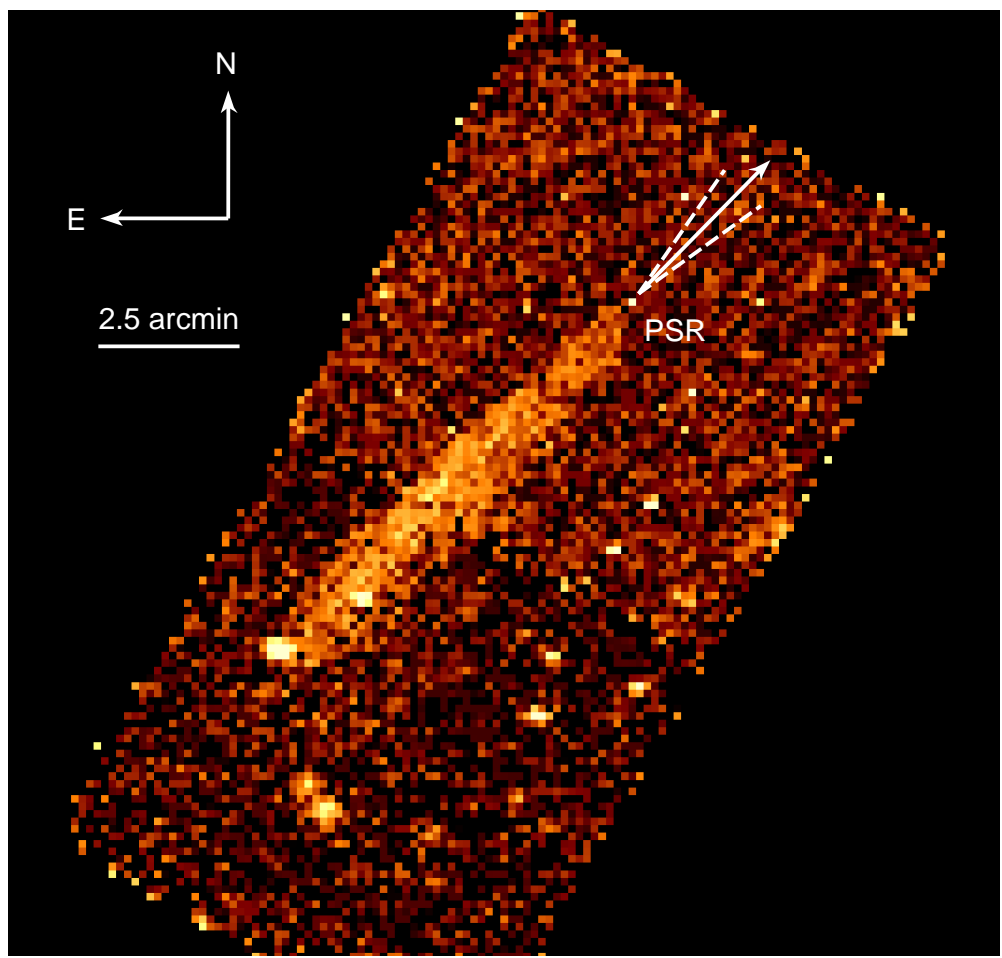}
\caption{The field of PSR J0357+3205 as seen by Chandra in the 0.3-8 keV
  energy range. Data from the two observations performed in 2009 have 
been used (77 ks exposure time). 
The image has been rebinned to a pixel size of 8$''$ in order
to ease visibility of faint diffuse structures. The large, elongated nebula is
apparent. The pulsar counterpart is marked, together with the direction of the
proper motion (arrow) and the $1\sigma$ errors (dashed lines).}
\end{figure}

\clearpage

\begin{figure}
\epsscale{.80}
\plotone{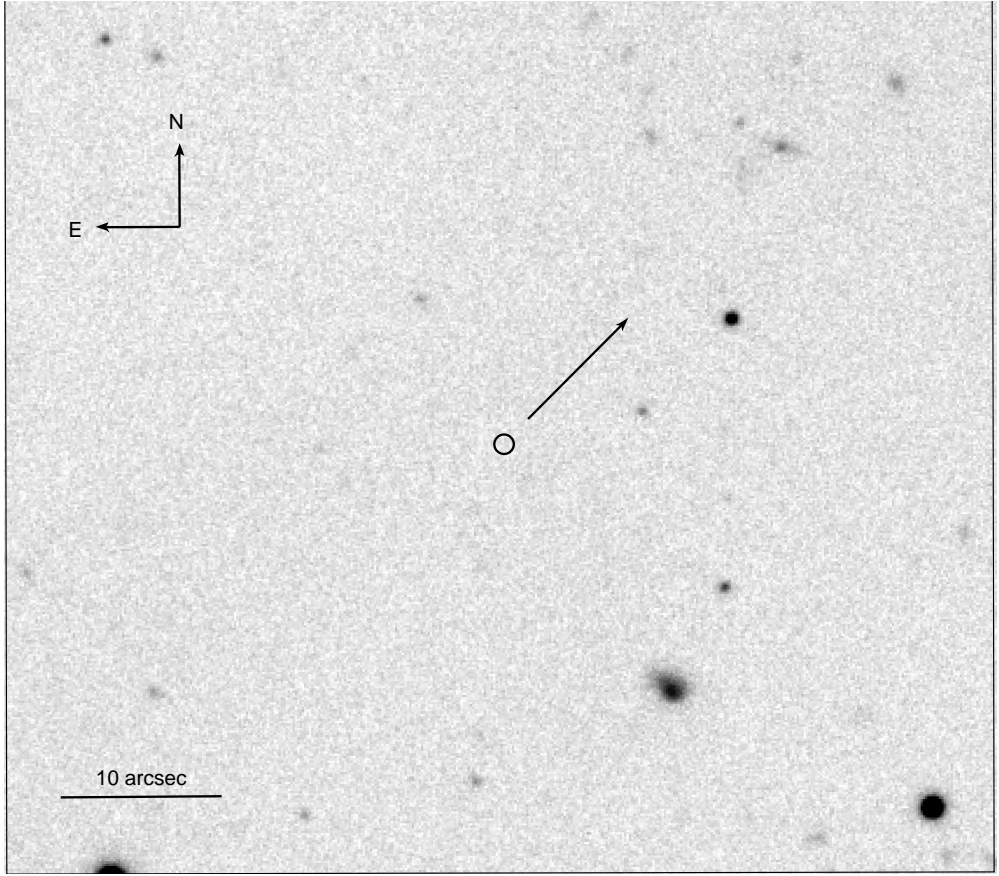}
\caption{Inner portion of the field of PSR J0357+3205 as seen by the
  Gemini/GMOS instrument in the $H_{\alpha}$ band. 
The position of the pulsar is marked by a circle ($0\farcs6$ radius). No diffuse emission is
  seen, related to the proper motion direction of the pulsar, marked by an
  arrow. 
The upper limit
  to an undetected bow-shock nebula is $5\times10^{-18}$ erg cm$^{-2}$
  s$^{-1}$ arcsec$^{-2}$.}
\end{figure}


\begin{thebibliography}{}

\bibitem[\protect\citeauthoryear{Abdo et al.}{2009a}]{abdo09a} Abdo, A.A., et
  al., 2009a, Science, 325, 840

\bibitem[\protect\citeauthoryear{Abdo et al.}{2009b}]{abdo09b} Abdo, A.A., et
  al., 2009b, ApJS, 183, 46

\bibitem[\protect\citeauthoryear{Atwood et al.}{2009}]{atwood09} Atwood, W.B.,
et al. 2009, ApJ, 697, 1071

\bibitem[\protect\citeauthoryear{Bandiera}{2008}]{bandiera08} Bandiera, R.,
2008, A\&A, 490, L3

\bibitem[\protect\citeauthoryear{Becker et al.}{2012}]{becker12} Becker, W.,
  Prinz, T., Winkler, P.F., Petre, R., 2012, ApJ 755, 141

\bibitem[\protect\citeauthoryear{Bertin \& Arnouts}{1996}]{bs96} Bertin E. \&
  Arnouts S., 1996, A\& A Suppl., 117, 393

\bibitem[\protect\citeauthoryear{Chatterjee \& Cordes}{2002}]{chatterjee02}
  Chatterjee, S., \& Cordes, J.M., 2002, ApJ, 575, 407

\bibitem[\protect\citeauthoryear{Cordes et al.}{1993}]{cordes93} Cordes, J.M.,
Romani, R.W., Lundgren, S.C., 1993, Nature 362, 133

\bibitem[\protect\citeauthoryear{De Luca et al.}{2011}]{deluca11} De Luca, A.,
Marelli, M., Mignani, R.P., et al., 2011, ApJ, 733, 104 

\bibitem[\protect\citeauthoryear{De Luca et al.}{2009}]{deluca09} de Luca, A.,
Caraveo, P.A., Esposito, P., Hurley, K., 2009, ApJ, 692, 158

\bibitem[\protect\citeauthoryear{de Zeeuw et al.}{1999}]{dezeeuw99} de Zeeuw,
P. T., Hoogerwerf, R., de Bruijne, J.H.J., et al., 1999, AJ, 117, 354

\bibitem[\protect\citeauthoryear{Dias et al.}{2002}]{dias02} Dias, W.S.,
Alessi, B.S., Moitinho, A., L´epine, J.R.D., 2002, A\&A, 389, 871

\bibitem[\protect\citeauthoryear{Gaensler \& Slane}{2004}]{gaensler06}
Gaensler, B.M. \& Slane, P.O., 2006, ARA\&A, 44, 17


\bibitem[\protect\citeauthoryear{Hui et al.}{2012}]{hui12} Hui, C.Y., 
 Huang, R.H.H., Trepl, L., et al., 2012, ApJ 747, 74

\bibitem[\protect\citeauthoryear{Hui \& Becker}{2007}]{hui07}Hui, C.Y. \&
Becker, W., 2007, A\&A, 467, 1209

\bibitem[\protect\citeauthoryear{Kaplan et al.}{2009}]{kaplan09} Kaplan, D.L.,
Chatterjee, S., Hales, C.A., Gaensler, B.M., Slane, P.O., 2009, AJ, 137, 354

\bibitem[\protect\citeauthoryear{Johnson \& Wang}{2010}]{johnson10} Johnson,
S.P. \& Wang, Q.D., 2010, MNRAS 408, 1216

\bibitem[\protect\citeauthoryear{Kargaltsev \& Pavlov}{2008}]{kargaltsev08}
  Kargaltsev, O., \& Pavlov, G. G. 2008a, in AIP Conf. Proc. 983, 40 Years of
  Pulsars: Millisecond Pulsars, Magnetars and More (Melville, NY: AIP), 171

\bibitem[\protect\citeauthoryear{Kulkarni \& Heiles}{1988}]{kulkarni88}
Kulkarni, S.R.\& Heiles, C., 1988, in ``Galactic and extragalactic radio
astronomy (2nd edition)'' (A89-40409 17-90). Berlin and New York,
Springer-Verlag, 1988, p.95

\bibitem[\protect\citeauthoryear{Lasker et al.}{2008}]{lasker08} Lasker, B.M.,
Lattanzi, M.G., McLean, B.J., 2008, AJ, 136, 735

\bibitem[\protect\citeauthoryear{Lattanzi et al.}{1997}]{lat97} Lattanzi
M.~G., Capetti A., \& Macchetto F.~D.\ 1997, A\&A, 318, 997

\bibitem[\protect\citeauthoryear{Maiz-Apellaniz}{2001}]{maiz01}
  Ma\'iz-Apell\'aniz, J., 2001, AJ, 121, 2737

\bibitem[\protect\citeauthoryear{Marelli et al.}{2012}]{marelli12} Marelli,
M., De Luca, A., Salvetti, D., et al., 2012, submitted to ApJ

\bibitem[\protect\citeauthoryear{Mason et al.}{1998}]{mason98} Mason, B.D.,
Henry, T.J, Hartkopf, W.I., William, I., ten Brummelaar, T., Soderblom, D.R.,
1998, AJ, 116, 2975

\bibitem[\protect\citeauthoryear{Mel\'nik et al.}{2009}]{melnik09} Mel\'nik, A. M. \& Dambis, A.K., 2009, MNRAS, 400, 518

\bibitem[\protect\citeauthoryear{Mignani et al.}{2012}]{mignani12} Mignani,
  R. P., Vande Putte, D., Cropper, M., et al., 2012, MNRAS, in press, arXiv:1212.3141

\bibitem[\protect\citeauthoryear{Motch et al.}{2007}]{motch07} Motch, C.,
Pires, A.M., Haberl, F., Schwope, A., 2007, Ap\&SS, 308, 217

\bibitem[\protect\citeauthoryear{Motch et al.}{2008}]{motch08} Motch, C.,
Pires, A.M., Haberl, F., Schwope, A., Zavlin, V.E., 2008, AIPC, 983, 354

\bibitem[\protect\citeauthoryear{Ray et al.}{2011}]{ray11} Ray, P.S., Kerr,
  M., Parent, D., Abdo, A.A., Guillemot, L., et al., 2011, ApJS 194, 17

\bibitem[\protect\citeauthoryear{Skrutskie et al.}{2006}]{skr06} Skrutskie M.~F., Cutri R.~M., Stiening R., et al.\ 200
6, AJ, 131, 1163

\bibitem[\protect\citeauthoryear{van Etten et al.}{2012}]{vanetten12} Van
Etten, A., Romani, R.W., Ng, C.-Y., 2012, ApJ, 755, 151

\end{thebibliography}
\end{document}